\documentclass[amsart,
 aps,
 prl,
 superscriptaddress,
 twocolumn]{revtex4-2} 

\usepackage{graphicx, verbatim,amsmath,amssymb, float}
\usepackage{color}

\begin{document}
 
\title{ Homogeneous crystallization in cyclically sheared frictionless
  grains }


\author{Weiwei Jin$^*$}
\affiliation{Department of Mechanical Engineering and Materials
  Science, Yale University, New Haven, Connecticut 06520, USA}
\author{Corey S.\ O'Hern}
\affiliation{Department of Mechanical Engineering and Materials
  Science, Yale University, New Haven, Connecticut 06520, USA}
\affiliation{Department of Physics, Yale University, New Haven,
  Connecticut 06520, USA}
\affiliation{Department of Applied Physics, Yale University, New
  Haven, Connecticut 06520, USA}
\author{Charles Radin}
\affiliation{Department of Mathematics, University of
Texas at Austin, Austin, Texas 78712, USA}
\author{Mark D.\ Shattuck$^{**}$}
\affiliation{Benjamin Levich Institute and Physics Department, The
  City College of New York, New York, New York 10031, USA}
\author{Harry L.\ Swinney}
\affiliation{Center for Nonlinear Dynamics and Department of Physics,
  University of Texas at Austin, Austin, Texas 78712, USA}

\begin{abstract} Many experiments over the past half century have
  shown that, for a range of protocols, granular materials compact
  under pressure and repeated small disturbances.  A recent experiment
  on cyclically sheared spherical grains showed significant compaction
  via homogeneous crystallization (Rietz {\it et al.}, 2018).  Here we
  present numerical simulations of frictionless, purely repulsive
  spheres undergoing cyclic simple shear with dissipative Newtonian
  dynamics at fixed vertical load. We show that for sufficiently small
  strain amplitudes, cyclic shear gives rise to homogeneous
  crystallization at a volume fraction $\phi = 0.646 \pm 0.001$. This
  result indicates that neither friction nor gravity is essential for
  homogeneous crystallization in driven granular media.
\end{abstract}

\maketitle


Loose granular materials will compact under a range of different
protocols of small repeated disturbances while under the influence of
gravity and/or external pressure. In the half century following the
pioneering work of J.D. Bernal \cite{Bernal59} and G.D.\ Scott
\cite{Scott60} around 1960,
the existence of a barrier to compaction has been confirmed for
many types of disturbances \cite{Knight1995_Richard2005,  Liber2013,
  Chen2006, Schroeter2005}, but not for cyclic shear for which the systems
compacted via wall-induced crystallization \cite{Scott64_Nicolas00}. Walls that
inhibit nucleation and precision measurements of the positions of the
spherical grains recently enabled Rietz {\it et al}.~ \cite{Rietz18} to observe
homogeneous crystallization in a granular material
undergoing cyclic shear.  

Using Newtonian overdamped dynamics, we numerically simulate purely repulsive spherical grains
undergoing cyclic simple shear strain  
dynamics to model granular crystallization
in the Rietz {\it et al}.\ experiment \cite{Rietz18}.  The
simulations allow us to tune the grain interactions, gravity, and
boundary conditions to understand the essential physics that gives
rise to homogeneous crystallization.  

Our simulations  reveal the  essential physical requirements
needed for crystallization in driven dissipative granular materials:  volume exclusion, system confinement,
and small disturbances that allow grain rearrangements. In particular, gravity, friction, and energy
conservation are not required to yield homogeneous crystallization.
The simulations use deterministic dynamics 
 in contrast with the probabilistic 
evolution equations used in classical nucleation theory~\cite{Abraham_Debenedetti} 
to model crystallization in atomic and molecular systems.  

From the simulations  we show that the volume fraction at the onset
of crystallization  becomes
independent of the shear amplitude for sufficiently small amplitudes,
 $A \le
0.05$  rad, consistent with Rietz {\it et al.}\ \cite{Rietz18},  who  used $A=0.01$ rad. The
volume fraction at the onset of crystallization is also within
the range of jamming volume fractions found for different packing
generation protocols for frictionless hard spheres~\cite{Chaudhuri}.

{\bf Methods.} We  simulate cyclic  simple shear of frictionless  monodisperse
spheres in three spatial dimensions using a discrete element method (see Supplemental Materials (SM), Sec. B \cite{SM}).
 Figure~\ref{fig_1}(a) shows the simulation shear cell, designed
to mimic the parallelepiped shear cell used in experiments
\cite{Scott64_Nicolas00} with
two sidewalls that tilt with respect to the vertical axis by a
variable angle $\theta$ in the $x$-$y$ plane,   and a bottom wall that oscillates horizontally and can move
vertically under an applied load.   The system is periodic in the
$z$-direction. The initial jammed packing is prepared via 
gravitational deposition by pouring the grains into a  static  container and allowing the dissipative
forces to remove the kinetic energy in the system.  A
subsystem, of size $(L+2d) \times (1.675)L \times L$ in the $x$-, $y$-, and $z$-directions, is
cut out from the initial packing, where $d$ is the grain diameter (see
Fig.\ref{fig_1}(b)).  Grains
with center positions less than $d$ away from all of the surfaces of the
subsystem except the two in the $z$ direction are then fixed to form the walls.
Wall-induced crystallization is suppressed since the grains forming the walls are
randomly positioned.  Simulations were run for cells with edge lengths
$L/d=13$, $15$, $17$, and $20$ containing, respectively, $N=5200$,
$7900$, $11400$, and $18200$ grains. The selection of the simulation
time step size is discussed in SM Sec. B.3  \cite{SM}. 

The grains are modeled as frictionless spheres interacting via the  pairwise, purely repulsive
linear spring force,
\begin{equation}
\vec{F}_{ij}=k d \delta_{ij} \Theta\left(\delta_{ij}\right) \hat{r}_{ij},
\label{eq:force}
\end{equation}
where $k$ is the spring constant (see SM Sec. B.1 \cite{SM}),
$\delta_{ij}=1-r_{ij}/d$ is the intergrain overlap, 
$r_{ij}$ is the
center-to-center separation between grains $i$ and $j$,
$\Theta\left(x\right)$ is the Heaviside step function, and
$\hat{r}_{ij}$ is the unit vector pointing from the center of grain $j$ to
the center of grain $i$.
The interaction between the fixed wall grains and the interior
grains follows the same force law as that between pairs of interior grains. 
Under gravity in  the  $y$-direction, Newton's equation of motion for
each interior grain $i$ is given by
\begin{equation}
m \vec{a}_i=\sum _{j=1}^{n_i} \vec{F}_{ij}-b (\vec{v}_i-\vec{v}_{\rm fluid})-m_i g \hat{y},
\label{eq:motiona}
\end{equation}
where $g\hat{y}$ is the acceleration due to gravity, $m_i$, $\vec{a}_i$, and $\vec{v}_i$ are the 
mass, acceleration, and velocity of grain $i$,
${n_i}$ is the number of grains overlapping grain $i$, and $b$ is the damping constant. 
The damping force arises from Stokes drag with the assumption of the 
existence of fluid in the cell, as in the experiment of Rietz {\it et al}.  We consider two forms for ${\vec v}_{\rm fluid}$ for the Stokes drag; see SM  Sec.\ B.2 \cite{SM}. 
The equations of motion are integrated using a modified velocity-Verlet scheme.
We are interested in simulating grains in the hard sphere limit; 
the relationship between the spring constant and the interparticle
overlap is discussed in SM Sec.\ B.1 \cite{SM}). 

To shear the packing, the bottom  wall oscillates in the
$x$-direction (see Fig.~\ref{fig_1} (a)) as the two sidewalls tilt
with an angle $\theta$ that evolves with time $t$ as
\begin{equation}
\theta\left(t\right) = A\sin (\omega t),
\label{eq:angle}
\end{equation}
where $A$ is the shear amplitude and $\omega$ is the angular frequency. 
We study shear amplitudes $A=0.03$, $0.04$, $0.05$, $0.0667$, $0.0833$, and $0.10$ rad. 
The oscillating bottom wall consists of $N_\text{b}$ grains and is subjected to a force $\vec{F}_{\text{b},j}$ from interior grains.
The motion of the bottom wall due to the applied pressure $P$ can be obtained by solving
\begin{equation}
m_\text{b}\vec{a}_\text{b}=\sum _{j=1}^{N_\text{b}}
\vec{F}_{\text{b},j}-b (\vec{v}_\text{b}-\vec{v}_{\rm fluid})-m_\text{b} g \hat{y}+\frac{L^2}{N_\text{b}}P \hat{y},
\label{eq:motionb}
\end{equation}
where $m_\text{b}$ is the total mass of grains in the bottom wall and
$\vec{a}_\text{b}$ and $\vec{v}_\text{b}$ are the acceleration and velocity of the
bottom wall,  which is constrained to move along the direction, $\hat{n}=(\sin \theta(t), \cos \theta(t),0)$, at time $t$.  
The values of the simulation parameters are listed in Table~\ref{table1}.

An alternative energy relaxation method was also considered, closer to
that used in the experiment of Rietz {\it et al}.~\cite{Rietz18}.
Simulations with amplitude $A=0.05$ rad were carried out with
the grains allowed to come to rest after each cycle of the
walls, instead of moving the walls and dissipating energy continuously.
Fig.\ S4 in Supplemental Material Sec.\ B.4 shows that the results for the 
fraction of crystallized grains versus volume fraction $\phi$ using
the two relaxation methods is similar until $\phi \approx 0.646$.

 \begin{table}
 \caption{Parameters used in the discrete element method simulations of cyclic simple shear.\label{table1}}
 \begin{tabular}{lc}
 \hline
 \hline
 Parameter & Value\\
 \hline
 Grain mass $m$ & $2\times10^{-5}$ kg\\
 Grain diameter $d$ & $3\times10^{-3}$ m\\
 Spring constant $k$ & $6.54\times 10^2$ N/m\\
 Damping constant $b$ & $1.28\times 10^{-3}$ kg/s\\
Pressure $P$ & $9.54\times 10^2$ Pa \\
 Angular frequency $\omega$ & $\pi$ rad/s \\
 Simulation steps per cycle & 45600\\
 Gravitational acceleration $g$ & 9.81 m/s$^{2}$\\  
 \hline
 \hline
 \end{tabular}
 \end{table}

\begin{figure}[htbp]
\centering
\includegraphics[width=0.47\textwidth]{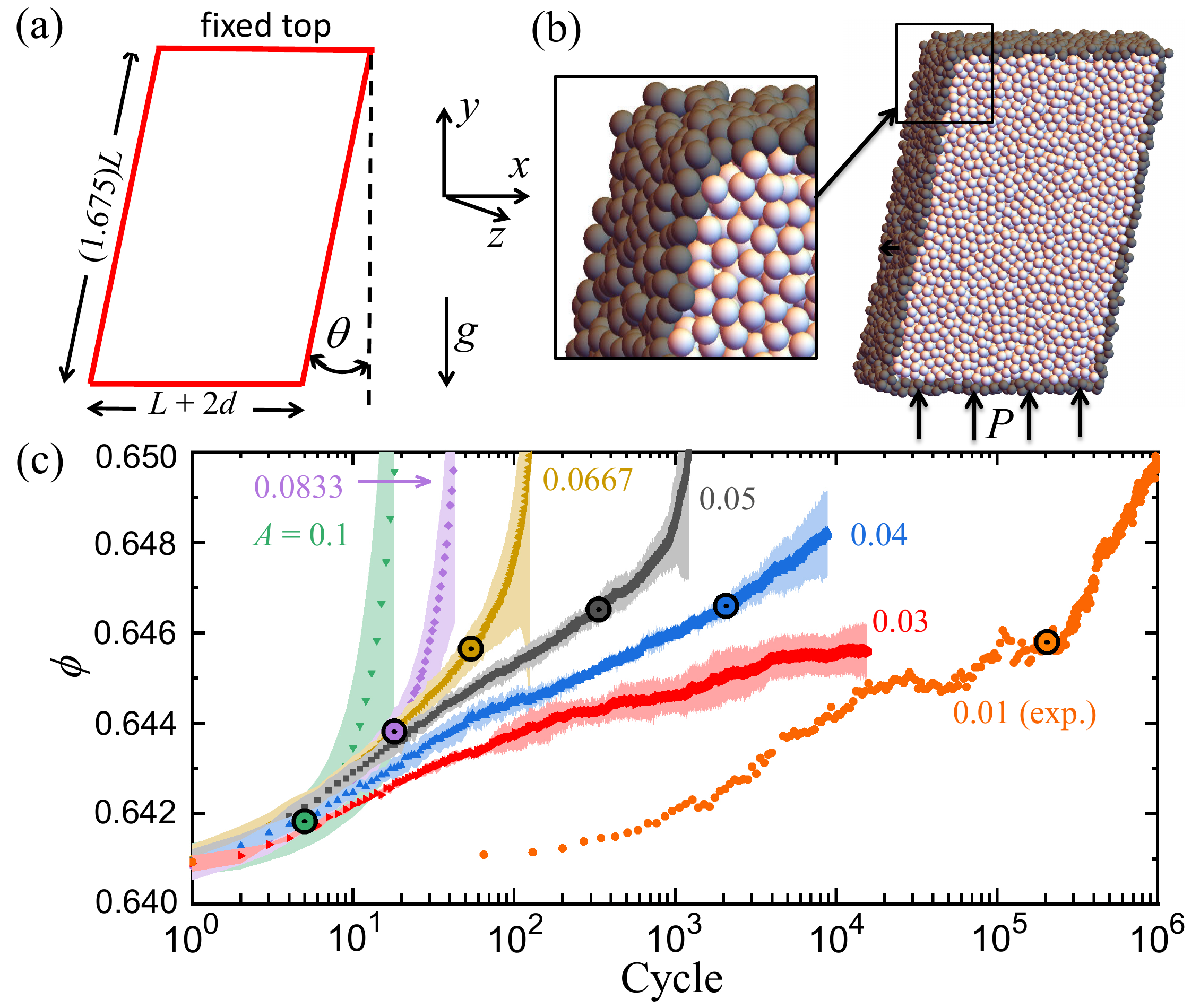}
\caption{(a) Cell used for simulations of frictionless spheres
  undergoing cyclic shear with $\theta=A \sin(\omega t)$. The cell
  dimensions are $(L+2d) \times (1.675)L \times L$, where $d$ is the
  diameter of the spheres; the boundary conditions in the $z$-direction are periodic.  Simulations were conducted for $L$ varying
  from $13d$ to $20d$. (b) The dark colored spheres on the cell
  boundary are fixed in disordered positions to inhibit
  crystallization. The light colored spheres in the cell interior move
  in the cyclically sheared cell in response to 
   interactions with one another, the dark spheres, and gravity. A constant pressure $P$ is
  applied on the bottom wall, which can move up and down. (c) The volume fraction $\phi$ as a function of cycle number in the simulations; the shading shows the standard deviation in simulations with different initial conditions. The results for $A=0.01$ rad (orange points) are from the experiments by 
Rietz {\it et al}.\ experiment \cite{Rietz18}.   
The  black circle with colored interior  on each curve shows the value of $\phi$ at the onset of rapid growth of crystallites in that system.}
\label{fig_1}
\end{figure}

\begin{figure}[htbp]
\centering
\includegraphics[width=0.4\textwidth]{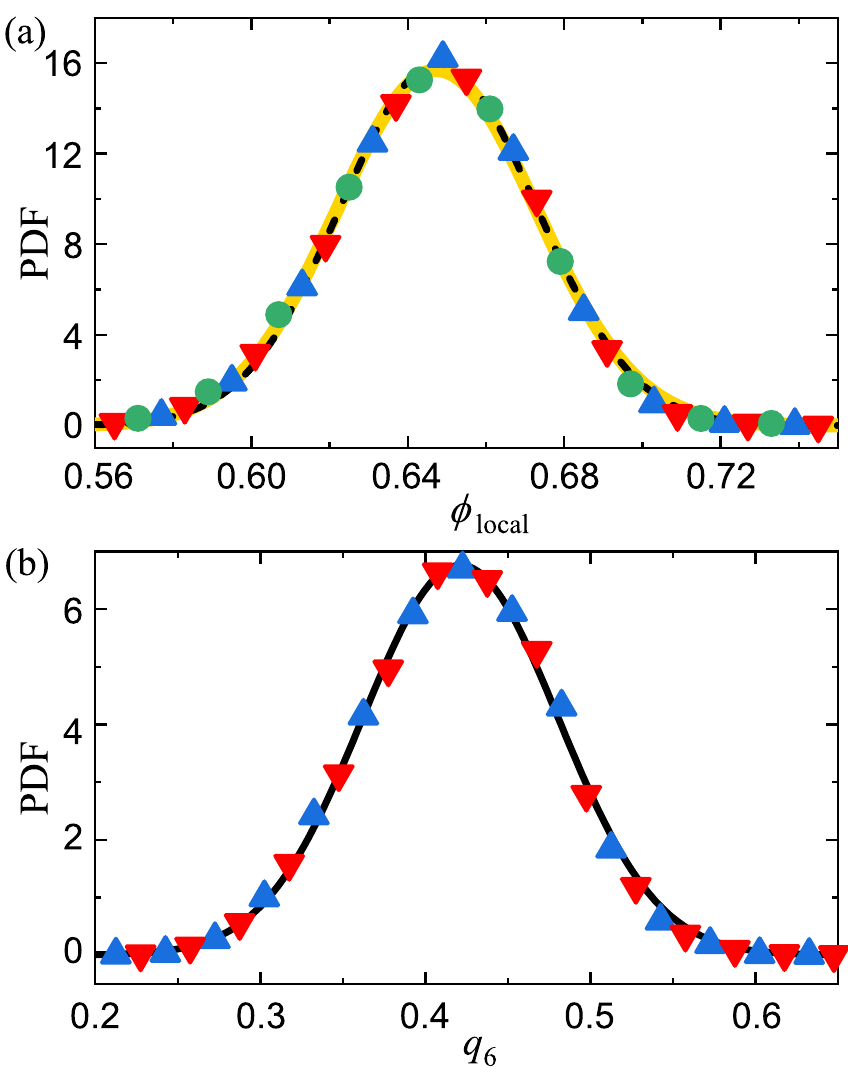}
\caption{(a)  The probability distribution function for the local
  volume fraction $\phi_{\rm local}$
 becomes {\it time-invariant}, as illustrated by the simulation results 
for shear amplitude $A=0.03$ rad at $n=10000$  shear cycles 
(blue upward triangles, $\phi=0.64581 \pm 0.00003$) and $15000$ shear cycles (red downward
triangles, $\phi=0.64593 \pm 0.00003$). The green circles  are results from the Rietz {\it et al}.\ experiment \cite{Rietz18}  
with  $A=0.01$ rad, averaged over shear cycles from  18849 to 67374, where the
volume fraction remained constant at  $0.6449 \pm 0.0001$.  
The black and yellow curves are respectively Gaussian fits
to the data for the simulations and experiment. 
 (b) The probability distribution function for the local
bond orientational order parameter $q_6$ 
is also time-invariant, as
illustrated  by  these results at
$n=10000$ (blue upward triangles) and $15000$ shear cycles (red downward triangles). The black curve is a Gaussian  fit.}
\label{fig_2}
\end{figure}

\begin{figure}[htbp]
\centering
\includegraphics[width=0.4\textwidth]{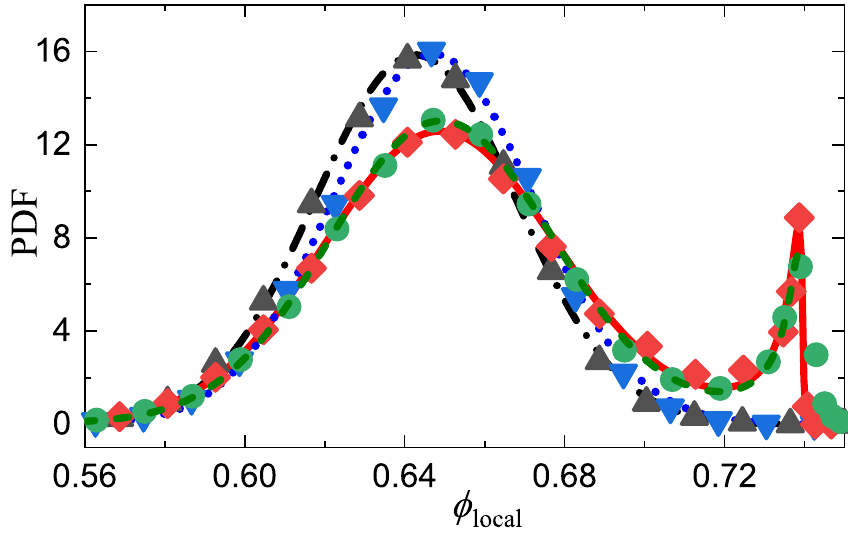}
\caption{Probability distribution function of the local volume fraction $\phi_{\rm local}$ for $\phi = 0.6406$
(black upward triangles), $ 0.6450$ (blue downward triangles), and $0.6538$ (red diamonds),
from simulations with $A=0.05$ rad. The black and blue curves  are  Gaussians. For $\phi = 0.6538$
(red diamonds), the probability distribution has a second peak at $\phi_{\rm local} \sim 0.74$,
which corresponds to crystallites with FCC and HCP symmetry. Experimental data  (green circles)  for $A=0.01$ rad at the same $\phi$  also possess a peak at 
$\phi_{\rm local} \sim 0.74$. The simulation and experimental data at $\phi= 0.6538$ can be fitted 
as the sum of a Gaussian $g(\phi_{\rm local})$ and an inverse Gaussian $f(\phi_{\rm local})$ (red and green curves),
$P(\phi)=a g(\phi_{\rm local}) + (1-a) f(\phi_c-\phi_{\rm local})$, where
$\phi_c =\pi/3\sqrt{2}$
(the few data points beyond $\phi_c$ are  
not included in the fit).}
\label{fig_3}
\end{figure}

\begin{figure}[!htbp]
\centering
\includegraphics[width=0.48\textwidth]{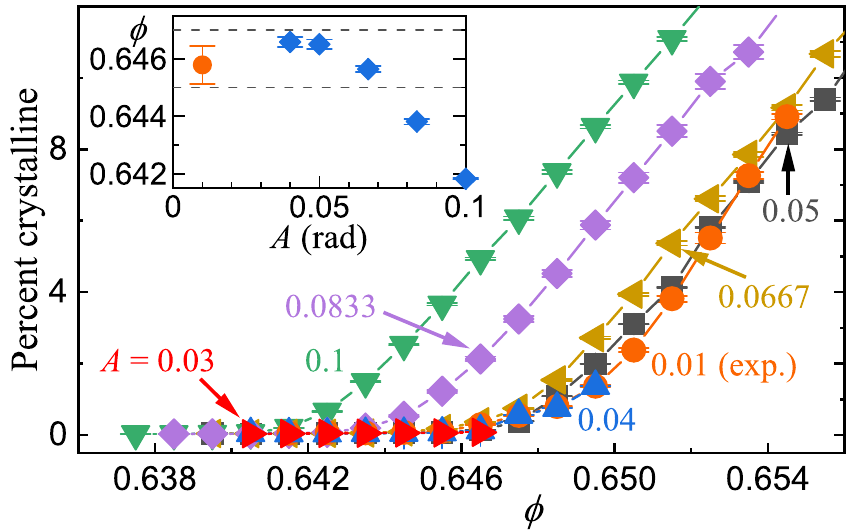}
\caption{The percentage of grains that are in crystallites as a function of
the volume fraction is shown for simulations with $A = 0.03$ to $0.10$ rad
and for  the experiment  with $A=0.01$ rad (orange circles) (Rietz {\it et al.}).
The results from the simulations with $A=0.05$ rad and the experiments with
$A=0.01$ rad are  remarkably similar.  The inset shows that the volume fraction at the
onset of crystallization approaches a well-defined value as the shear
amplitude decreases (blue diamonds: simulations; orange circle: experiments).
The volume fraction at onset  is defined  to correspond to the emergence
of a positive growth rate for the largest crystallite (10 grains
for $A=0.01$, $0.04$, and $0.05$; 13 for $A=0.0667$; 16 for $A=0.0833$;
and 20 for $A=0.10$).}  
\label{fig_4}
\end{figure}

{\bf Results.} Intuitively, the global volume fraction $\phi$ of a collection of grains
in a container is the sum of the volumes of the grains divided by the
volume of the container. If one partitions the volume of the container
by the Voronoi cells of the grains,  one has a useful local version, the
local volume fraction $\phi_{\rm local}$ for that Voronoi cell or grain. Our grains are
slightly deformable so there is some choice in how to assign a volume to
them.  We are  interested in
modeling asymptotically hard grains, as in the experiment Rietz {\it
  et al}, 
so we define the volume of a
grain to be  $(\pi d^3/6)(1-\delta_{\rm max})^3$, where
$\delta_{\rm max}$ is the maximum of the `overlap' volume of that sphere with
its neighbors (see Supplemental Material Sec.\ B.1). The global volume fraction of a collection of grains is
then the mean of the local volume fractions of all the `free' grains,
that is, the grains which are fixed as part of the boundary are not
counted.  

The results for the global 
 volume fraction $\phi$ obtained in our
simulations for shear amplitudes $A=0.03$, $0.04$, $0.05$, $0.068$, $0.083$,
and $0.10$ rad are shown in Fig.~\ref{fig_1}(c), along with results from the
experiment of  Rietz {\it et al}.\  \cite{Rietz18} for $A=0.01$ rad.  The volume fraction 
increases in a similar way for each shear amplitude for the first few
shear cycles (i.e., $n<10$), while subsequently the increase in volume fraction 
is more rapid for larger shear amplitudes, as seen already by
Scott experimentally \cite{Scott64_Nicolas00}.

At low shear amplitudes and for sufficiently small cycle numbers,
the system remains disordered.  Figure~\ref{fig_2}(a) shows, for $A=0.03$ rad
and several cycle numbers, probability distribution functions for the local volume fraction
$\phi_{\rm local}$. These probability distributions become time-invariant after  $n \sim
10000$  shear cycles. 
Figure~\ref{fig_2}(b) shows time invariance 
also in the probability distribution functions for
 the  local bond orientational order $q_6$~\cite{Steinhardt83}, which we use to identify
crystallization; see Supplemental Material Sec.\ A for background.
Similarly, the experiment with $A=0.01$ rad reached a persistent state at
$1.1 \times 10^5$ shear cycles (see Fig.~\ref{fig_1}(c)). 

The simulations with amplitudes $A \geq
0.05$ rad do not give rise to a persistent state, as can be seen in Fig.~\ref{fig_1}(c).
Further, Fig.~\ref{fig_3} shows that for $A=0.05$ the
probability distribution function for the local density starts as a
Gaussian with a peak near $\phi_{\rm local}=0.65$.  With an increasing
number of shear cycles, the height of this peak decreases and its width
increases. With continued shearing, a second peak emerges at
$\phi_{\rm local} \approx 0.74$  in both the experiment with $A=0.01$ rad and
in the simulations with $A=0.05$ rad. (Thus, the persistent state in the
simulations at $A=0.03$ rad begins to disappear for shear amplitudes 
in the range $0.03\, {\rm rad} < A < 0.05\, {\rm rad}$ and is absent for  larger  $A$.) 
The second peak in the probability
distribution at 
 $\phi_{\rm local} \approx 0.74$ corresponds to the formation
of HCP and FCC crystallites.

 Figure~\ref{fig_4} presents results for the fraction of the system that is
 crystalline in simulations with different amplitudes $A$.  (The identification of crystalline
 grains is discussed in SM Sec. A  \cite{SM}.) The
 similarity between the results from the simulation for $A=0.05$ rad and the experiment for $A=0.01$ rad is remarkable considering
the many differences between the simulation model and the laboratory
experiment.  For $A > 0.05$ rad (i.e., $0.067$, $0.083$, and $0.10$ rad), the crystallization onset 
occurs  for a smaller volume fraction than the onset observed in the experiment. 

We have considered several methods for obtaining an accurate estimate for
the volume fraction at the onset of crystallization in
a system satisfying the limits of small shear amplitude and large system size.
The method used for the inset of Fig.~\ref{fig_4}   
is to determine the critical
cluster size  for each shear amplitude $A$ and  identify the
associated value of the volume fraction; see Supplemental Material Sec.\ C for a comparison with
an alternative method for determining the volume fraction at the onset of
crystallization. Both methods are in 
quantitative agreement, yielding the results in the inset of Fig.~\ref{fig_4}.

{\bf Discussion.} The results obtained from the simulations of the
deterministic equations of motion for frictionless 
monodisperse spheres in a container
undergoing periodic shear reveal the onset of homogeneous
crystallization that depends on the shear amplitude $A$ 
and on the number of grains $N$ in the system. 
However, the simulations show that there is a well-defined volume
fraction, $\phi = 0.646 \pm 0.001$, above which homogeneous crystallization 
occurs in the limit of small amplitude, 
$A \leq 0.05$ rad (see inset of Fig. 4(a)) and large system size, $N \geq
18200$. 
This characteristic volume fraction $\phi = 0.646$ agrees well 
with the experimental value found in Rietz {\it et al}.~\cite{Rietz18}, where 
$A=0.01$ rad and $N=49400$.  

Note that there are multiple differences between the experiment by
Rietz {\it et al.}\ and the simulations, including the form of the
dissipation, which in the experiment arises from Stokes drag on the
motion of the grains relative to the viscous
refractive-index-matching fluid surrounding the grains, from the
inelasticity of the grains, and from the static and dynamic friction
between grains in contact.  In contrast, in the simulations only
Stokes dissipation is included; 
 the static and kinetic friction
coefficients are zero. Still, the results for the onset of crystallization in
the simulations and the Rietz {\it et al.}\  experiment 
agree in the limits of hard-sphere interactions, low driving amplitude, and large
system size.

The {\it random close packed} (RCP) limit, which has been discussed in
more than one thousand papers in the past half century, has never been
unambiguously defined and may well be different for different physical
compaction protocols~\cite{Torquato_Radin_Makse}. However, for cyclic simple shear (in the quasistatic
limit), we find a well-defined value of the global volume fraction,
$0.646 \pm 0.001$ at the onset of crystallization.  This value, which
is in the reported range for RCP, 0.62-0.66,  
is robust to friction, gravity, and amplitude changes in the small amplitude limit. To avoid
confusion we refer to the value we obtain as the 
`crystallization volume fraction’ for asymptotically small shear
amplitudes.

There are compaction protocols, such as cyclic shear coupled with
horizontal and vertical shaking, for which the particles or grains possess
significant kinetic energy that is slowly drained during the
compaction process. For such
protocols, it is possible as noted above, that the volume fraction for the onset of
crystallization may  occur at a different value than the one found
here.  One method to investigate this question is to carry out
simulations of cyclic shear for underdamped, frictionless grains with
weak damping, in addition to weak horizontal vibrations.  
Then the onset of crystallization could be determined 
 as a function of the damping coefficient~\cite{shattuck} as well as 
in the limit of small shear and vibration amplitudes. 

Physical nongranular crystallization experiments
and their modeling, in
particular the homogeneous freezing of molecular fluids 
\cite{Abraham_Debenedetti},
should be compared with the
homogeneous crystallization in the present simulations and in Rietz
{\it et al.}~\cite{Rietz18} and \cite{Swinney19}. In particular,
persistent states play an important role in classical nucleation theory
but seem to play a marginal role in these simulations and
experiment, disappearing with growing shear amplitude.

{\bf Acknowledgements} We acknowledge support from NSF Grant Nos.
CBET-2002782 (C.O.), CBET-2002797 (M.S.), and DMR-1119826 (W.J.) and
from the Army Research Laboratory under Grant No. W911NF-17-1-0164
(C.O.). This work was also supported by the High Performance Computing
facilities operated by Yale's Center for Research Computing and
computing resources provided by the Army Research Laboratory Defense
University Research Instrumentation Program Grant No. W911NF1810252.\\

$^{*}$microwei.jin@gmail.com\                
$^{**}$shattuck@ccny.cuny.edu \\

\end{document}